\documentclass[preprint,superscriptaddress,showpacs,preprintnumbers,amsmath,amssymb,aps]{revtex4}
\usepackage{graphicx}% Include figure files
\usepackage{dcolumn}% Align table columns on decimal point
\usepackage{bm}% bold math
\usepackage{color}
\usepackage{enumerate}
\begin{document}

\title{Identification of Seismic Electric Signals upon significant data loss}

\author{P. A. Varotsos}
\affiliation{Section of Solid State Physics and Solid Earth Physics
Institute, Department of Physics, School of Science, National and Kapodistrian University of Athens,
Panepistimiopolis, Zografos 157 84, Athens, Greece}
\author{N. V. Sarlis}\email{nsarlis@phys.uoa.gr}
\affiliation{Section of Solid State Physics and Solid Earth Physics
Institute, Department of Physics, School of Science, National and Kapodistrian University of Athens,
Panepistimiopolis, Zografos 157 84, Athens, Greece}
\author{E. S. Skordas}
\affiliation{Section of Solid State Physics and Solid Earth Physics
Institute, Department of Physics, School of Science, National and Kapodistrian University of Athens,
Panepistimiopolis, Zografos 157 84, Athens, Greece}

\begin{abstract}
When monitoring geophysical parameters, data from segments that are contaminated by noise may have to be abandoned. This is the case, for example, in the geoelectrical field measurements at some sites in Japan, where high noise -due mainly to leakage currents from DC driven trains- prevails almost during 70\% 
of the 24 hour operational time. We show that even in such a case, the identification of seismic electric signals (SES), which are long-range correlated signals, may be possible, if the remaining noise free data are analyzed in natural time along with detrended fluctuation analysis (DFA). 

\end{abstract}

% Keywords required only for MST, PB, PMB, PM, JOA, JOB? %
\maketitle
\section{Introduction}

In many cases of geophysical and/or geological interest, it happens that for substantial parts of the time of data collection, high noise prevents any attempt for extracting a useful signal. Data for such time segments are removed from further analysis. The appearance of such a noise may be periodic as in the case treated in the present work. It is the objective of this paper to examine whether the remaining data allow the identification of long-range temporal correlations. 

The present study was motivated from the results of geoelectrical measurements in Japan aiming at the detection of Seismic Electric Signals (SES), which are low-frequency ($\leq$1 Hz) variations of the electric field of the earth that precede earthquakes, e.g., see \cite{VAR84A,VAR84B,VAR96BACTA} and references therein. SES sometime appears as a single signal lasting for minutes but often many SES (hereafter called “pulses” as needed) keep appearing during certain length of time, which may be as long as a few days or more. Such a case is called SES activity. It has been shown that the SESs in a SES activity have long range temporal correlations characteristic to critical phenomena\cite{NAT02,NAT02A} in accordance to an SES generation model based on solid state aspects discussed below\cite{VAR93A}, see also \cite{VAR08438} and references therein. The measurements in Japan have detected clear SES either at noise-free measuring sites or at noisy stations when the SES happened to occur at midnight, i.e., when the noise level was low\cite{UYE00,UYE02}. 
The major difficulty at many sites is the contamination of records by high noise due to leakage currents from DC driven trains and other artificial sources, against which some countermeasure such as independent component analysis to extract signals has been attempted (e.g., see \cite{ORI09}). The low noise time occurs from 00:00 to 06:00 and from 22:00 to 24:00 local time (LT) when nearby DC driven trains cease service, i.e., almost only 30\% of the 24 hours. Thus, the question arises whether it is still possible to identify SES upon removing the noisy data segments lasting for the period 06:00 to 22:00 every day. The answer to this question is attempted in this paper for the case when the duration of SES activity is much longer compared to those of individual pulses, i.e., a few days to a few weeks or even more, although admittedly long lasting SES activity is rather seldom, limiting the applicability of the results described below.

The key point in the present work is the use of the following two modern methods: The natural time analysis of the remaining data and the detrended fluctuation analysis (DFA). The present question differs from the one in which we investigated\cite{SKO10} the effect of the random in time removal of data segments of fixed length on the scaling properties of SES activities. It also differs from the case in which the lengths of the lost or removed data segments are random and may follow a certain type of distribution\cite{MA10}.

We now briefly describe the time series analysis in natural time $\chi$, which
is a new time domain\cite{NAT02,NAT02A,NAT03A,NAT03B,NEWBOOK}. In a time series
comprising $N$ events, the natural time $\chi_k = k/N$ serves as
an index for the occurrence of the $k$-th event. The evolution of
the pair ($\chi_k, Q_k$) is studied, where $Q_k$ is a quantity
proportional to the energy released in the $k$-th event.
 For dichotomous
signals, which is frequently the case of SES activities, the
quantity $Q_k$ {can be replaced by}  the duration
of the $k$-th pulse. By defining $p_k=Q_{k}/\sum_{n=1}^{N}Q_{n}$,
we have found that the variance $\kappa_1=\langle \chi^2 \rangle
-\langle \chi \rangle ^2$, where $\langle f( \chi) \rangle =
\sum_{k=1}^N p_k f(\chi_k )$,
 of the natural time $\chi$ with respect to the distribution $p_k$
 may be used for idenifying {\em criticality}, and hence the SES
 activities.  {More specifically}, the following relation
 should hold {for SES activities}
\begin{equation} \label{eqk1}
 \kappa_1\approx 0.070.
 \end{equation}
Beyond the condition of Eq. (1), we have shown that the SES activities, when analyzed in natural time, exhibit infinitely ranged temporal correlations and obey the conditions\cite{NAT05B,NAT06A,NAT06B}:
\begin{equation}\label{eq2}
    S, S_- < S_u.
\end{equation}
where $S$ is the entropy $S$ in natural time defined as\cite{NAT03B} 
 $S \equiv  \langle \chi \ln \chi \rangle - \langle \chi \rangle
\ln \langle \chi
\rangle$  and $S_{-}$  is the entropy obtained upon time reversal. Equation (2) states that both $S$ and $S_-$  are smaller than the value $S_u (=\ln 2 /2-1/4\approx
0.0966$) of a ``uniform'' (u) distribution, e.g. when all $p_k$
are equal.

The fact that SES activities exhibit {\em critical dynamics}, is believed as mentioned to be related to their generation mechanism (see \cite{VAR93A} and references therein). In the focal area of an impending earthquake (EQ hereafter), which contains ionic materials, the stress gradually increases. In ionic solids a number of extrinsic defects are always formed because they contain aliovalent impurities. These extrinsic defects are attracted by the nearby impurities and hence form electric dipoles the orientation of which can change through defect migration. When the stress (pressure) $\sigma$ reaches a critical value $\sigma_{cr}$, a {\em cooperative} orientation of these dipoles occurs generating SES. 

We now summarize the detrended fluctuation analysis DFA\cite{PEN94,TAQ95} which is a novel method that has been developed to address the problem of accurately quantifying long range correlations in non-stationary fluctuating signals. It has been applied to diverse fields ranging from DNA\cite{PEN93,STA99}, to meteorology\cite{IVAN99}, and economics \cite{VAN97,IVA04B}. DFA is, in short, a modified root-mean-square (rms) analysis of a random walk.   In principle, it estimates the deviations from the local trends $y_s(n)$  of a non-stationary time series of length $N$ piecewise by dividing it into small segments with length $s$ and compute the Fluctuation function $F(s)$, which is the variance of $y_s(n)$:
\begin{equation}\label{fluctfun}
{F(s)=\sqrt{\frac{1}{N}
\sum_{n=1}^N[{y}_s(n)]^2}}
\end{equation}
$F(s)$  corresponds to the trend-eliminated root mean square displacement of the random walker. Then, the above computation is repeated for a broad number of scales $s$ to provide a relationship between $F(s)$ and $s$.

When a power-law relation between $F(s)$ and $s$, i.e.,
\begin{equation}\label{fsmes}
{F(s) \sim s^{\alpha}}
\end{equation}
is found, it indicates the presence of scaling-invariant (fractal) behavior embedded in the fluctuations of the signal\cite{PEN94,TAQ95}. 
The fluctuations can be characterized by the scaling exponent $\alpha$, a self-similarity parameter: If $\alpha$=0.5, there are no correlations in the data and the signal is uncorrelated (white noise); the case $\alpha<$  0.5 corresponds to anti-correlations, meaning that large values are most likely to be followed by small values and vice versa. If $\alpha>$  0.5, there are long-range correlations, which are stronger\cite{BAS08} for higher $\alpha$. Note that $\alpha>$  1 indicates a non-stationary local average of the data and the value $\alpha$ =1.5 indicates Brownian motion (integrated white noise).

{For stationary signals with long-range power-law
correlations the value of the scaling exponent $\alpha$ is
interconnected with the exponent $\beta$ characterizing the power
spectrum $S(f) \sim f^{- \beta}$ ($f$=frequency)
through\cite{PEN93}}

\begin{equation}\label{alphabita}
{\beta = 2 \alpha - 1}
\end{equation}
 When employing natural time, DFA seems to
distinguish\cite{NAT03A} SES activities from some artificial
{noise} because, for the SES activities the
$\alpha$-values lie approximately in the range

\begin{equation}\label{EQ7}
0.9 \leq \alpha \leq 1.0{,}
\end{equation}
 {while} for artificial
{noise (caused by man-made sources) investigated in
Greece\cite{NAT03A,NAT03B}} the $\alpha$-values are markedly
smaller, i.e., $\alpha$=0.65-0.8. {In other words,
the artificial noise recorded in Greece, which at the most lasts
24 hours, may have long-range correlations, e.g. $\alpha \approx
0.75$ (see Fig.9 of \cite{NAT03B}), but none of  several artificial
noises studied  was found to exhibit infinitely ranged
long-range correlations (i.e., having $\alpha$-value close to
unity).}

\section{Data analysis and results}\label{sec2}

Let us suppose that we have a long time series of data $s(i)$ (shown in red in the example of Fig.1), with a duration appreciably larger than 24 hours for instance, and we are forced to remove the same segment of these daily data. The portion of the 24 hour data that remain will be hereafter labeled $p_r$ and the number of data corresponding to one period, say 24 hours, $T$. Thus, every $T$ samples, $(1-p_r)T$ of them (belonging to the shaded parts of Fig. 1) are removed. The remaining segments (blue in Fig. 1) are concatenated to form the new time series $c(i)$ which is subsequently read in natural time. We now impose the following conditions (7) and (8) on $c(i)$ for classifying the signal as SES activity. The condition (7) comes from the relation (6) after considering the reasonable experimental error:
\begin{equation}
\label{dfacon} 0.85 \leq \alpha \leq 1.10
\end{equation}
The condition (8) comes from Eqs. (1) and (2) also by considering the reasonable experimental error in $\kappa_1$:
\begin{equation}
\label{natcon} |\kappa_1-0.07|\leq 0.01, S\leq S_u, S_-\leq S_u
\end{equation}

In the following subsections, in order to solve our problem, synthetic signals will be produced and analyzed whether they obey conditions (7) and/or (8) using a Monte Carlo comprising $10^3$ realizations.
The Monte Carlo procedure has been used to ``average'' over the possible realizations of the synthetic SES activities and noises that will be discussed later in subsections \ref{2.1} and \ref{2.2} as well as the fact that both types of electric signals may start any time of the day. Thus, one should randomly select an integer $i_{init}$ from 1 up to $T$, and keep in $c(i)$ the samples $s(i_{init})$ to $s(i_{init}+p_rT-1)$ of $s(i)$, i.e., we keep $p_rT$ samples in total. The next segment to be kept in $c(i)$ is $(1-p_r)T$ samples after $s(i_{init}+p_rT-1)$, starting from $s[i_{init}+p_rT-1+(1-p_r)T+1= i_{init}+T]$ up to $s(i_{init}+T+p_rT-1)$ and so on (see the blue lines in Fig.1). This way we periodically remove $(1-p_r)T$ samples and keep $p_rT$ every $T$ samples from the original signal $s(i)$. This Monte Carlo simulation allows us to evaluate the probability to identify the original signal as an SES activity.

The probability that the condition (7) is satisfied will be hereafter labelled $p_1$. By the same token, the probability to satisfy the conditions (8) is designated by $p_2$. Finally, the probability to obey either condition (7) or conditions (8) will be labelled $p_3$.  Upon considering the number of the Monte Carlo realizations ($M=10^3$), a plausible estimation error (3STD/$\sqrt{M}$) at the most around 5\% is expected  (cf. 1/$\sqrt{10^3} \approx$0.032, and STD stands for the standard deviation of the quantity calculated by Monte Carlo, e.g., see Ref. \cite{MC}).

Since we are interested in the low cultural noise night-window, we hereafter focus on the $p_r$ values varying from $p_r =$ 0.2 to roughly $p_r =$ 0.3. The following two subsections summarize the results obtained for synthetic long duration SES activities and synthetic noise, respectively. The first subsection concludes that when using $p_3$ the remaining data are highly probable to enable the identification of an SES activity. This means that the use of either condition (7) or conditions (8), i.e. $p_3$, is better to be used for the classification of a signal as a SES activity. Thus, such a probability can be thought as the conditional probability p(SES$|$SES), i.e., the probability to identify a true SES activity as such, after periodic data loss (cf. the complementary probability to miss a true SES is labelled $p_{miss}\equiv $ 1-p(SES$|$SES)). The second subsection deals with another conditional probability p(SES$|$noise), labelled $p_{noise}$ (Fig.2),  to erroneously identify a noise as SES activity when using conditions (7) and (8). Finally, results from a real world data set are provided in the third subsection.

\subsection{Results obtained from synthetic long duration SES activities}\label{2.1}

In order to construct synthetic long duration SES activities, one has to make use of a model for both the time-series of the pulse durations $Q_k$ and the appropriate waiting times $W_k$, i.e., the times elapsed between consecutive emissions of SES pulses. Since for actual data (e.g. see Ref. \cite{NAT03A}) $Q_k$ and $W_k$ exhibit different behavior when studied by DFA, we follow below different methods to generate the corresponding time series: 
We start with the results of the DFA of the pulse durations $Q_k$:  Its DFA analysis reveals\cite{NAT03A} an exponent $\alpha \approx$1, i.e., 
$Q_k$ exhibit very strong long-range correlations (infinitely ranged long-range correlations). This led\cite{NAT06A} to study in natural time the one sided segments of a fractional Gaussian noise (fGn). fGn, which is stationary and Gaussian with zero mean, corresponds to the time series of the increments of a fractional Brownian motion (fBm) $B_H(t)$ with self-similarity index $H$ \cite{MAN68}; the latter is the Gaussian generalization of the Brownian motion –random walk- in the sense that its root mean square displacement is proportional to the $H$-th power of the time lag, i.e.,  $\sqrt{\langle [B_H(t)-B_H(s)]^2 \rangle} \propto |t-s|^H$  (cf. for $H=0.5$ we simply recover the behavior of a random walk).  Thus, for $H=0.5$ fGn is just white noise whereas for $H>0.5$ it corresponds to DFA exponents $\alpha \approx H$.  Here, we model the durations $Q_k$ as the one sided segments of a fractional Gaussian noise with $H\approx 1$, because Varotsos et al \cite{NAT06A} have shown that in this case both conditions (7) and (8) are satisfied. We now turn to the time series of the waiting times $W_k$ and consider that for actual data their DFA exponent $\alpha$ scatters around 
$\alpha \approx$0.5 (e.g. see Table III of Varotsos et al. \cite{NAT03A}), which reflects that there are almost no temporal correlations or at least the absence of significant long-range time correlations. Additionally, in the Supplemental Material of Varotsos et al. \cite{NAT03B}, we showed that the waiting times' statistics cannot be fully accounted for by an exponential distribution, thus they cannot be fully characterized as simply random (Poissonian). Moreover, their distributions for various SES activities (e.g. see Table II of the Supplemental Material of Varotsos et al.\cite{NAT03B}) do not seem to share obviously common properties. This may be understood in the frame of the aforementioned SES generation mechanism (see Ref. \cite{VAR93A} and references therein) during the preparation stage of a strong EQ:  A simultaneous achievement of the stimulating (critical) stress $\sigma_{cr}$ at all points in the stressed EQ preparation volume is not intuitively expected\cite{VAR91}, see also p.258 of Ref.\cite{NEWBOOK}. When the condition $\sigma = \sigma_{cr}$ is fulfilled in a sub-volume of the EQ preparation volume, an SES pulse is emitted. Thus, when this occurs consecutively in various sub-volumes, a series of SES pulses is emitted which constitutes an SES activity. In other words, the ``points'' obeying the condition $\sigma = \sigma_{cr}$ should lie on a ``surface'' A, let us call it critical stress front, which may be very complicated in view of the existing inhomogeneities. This front sweeps through the preparatory volume with highly stochastic (stick-slip like) movement, the details of which (e.g., the ``stick'' intervals) determine the waiting times $W_k$ and may depend on the specific preparation zone and the time to the mainshock. Such a complex behavior is difficult to model, but in the following we will assume as first approximation that the waiting times $W_k$ between consecutive SES pulses are independent and identically distributed (iid) random variables from an exponential distribution (in simple words, this means that $W_k$ come from an unbiased -ideal- exponential random number generator). This assumption will be also applied to the case of $W_k$ for the synthetic noise treated in the next subsection.

 Along these lines,
%assuming such a modelling for the long duration SES activities,
 we generate synthetic dichotomous signals $d(t)$ that
resemble long duration SES activities by the procedure described in detail below:
\begin{enumerate}[(i)]
\item We first generate the durations of the
pulses $T_{on}(k)\equiv Q_k$, $k=0, 1, 2,\ldots, N$ with $N \geq
200$, that come from the one sided segments of a fractional
Gaussian noise with $H=1$ and satisfy both conditions
(\ref{dfacon}) and (\ref{natcon}) {(cf. the method
to produce such segments, has been described in detail in
 Ref.\cite{NAT06A}  and can be also visualized in Fig.5 of the Supplemental
Material of that reference).} 

\item We then generate the waiting times
between consecutive pulses, i.e, $T_{off}(k)\equiv W_k$, which are
independent exponentially distributed random variables with
average value $\tau_{off}=\tau_{on} /\lambda$, where $\tau_{on}$
is the average duration of pulses (i.e., $\tau_{on} \equiv \langle
T_{on}(k) \rangle$) and $\lambda$ an arbitrary dimensionless parameter corresponding to the ratio of the probabilities to find the 
system active over the probability to find the system inactive. Since we investigate the case of long duration SES activities, the aforementioned critical stress front A, obeying the condition $\sigma = \sigma_{cr}$, is naturally expected to propagate ``slowly''. This should correspond to long ``stick'' intervals $W_k$ and hence increased probability to find the system inactive. For these reasons, a typical value corresponding 
to ten times higher probability to find the system inactive than active, i.e., $\lambda$=0.1, has been considered.

\item We
then construct $d(t)$ by assigning to it the value of 1 if the
system is active, i.e., during the emission of a pulse, at time
$t$, or 0 otherwise. Namely, we have $d(t)=0$ for $t <
T_{off}(0)$, $d(t)=1$ for $t\in
[T_{off}(0),T_{off}(0)+T_{on}(1)]$, then $d(t)=0$ again for $t\in
(T_{off}(0)+T_{on}(1),T_{off}(0)+T_{on}(1)+T_{off}(1))$  and
$d(t)=1$ for $t \in
[T_{off}(0)+T_{on}(1)+T_{off}(1),T_{off}(0)+T_{on}(1)+T_{off}(1)+T_{on}(2)]$
and so on.
\end{enumerate}

The time-series $d(t)$ generated this way was then periodically cut, following the Monte Carlo procedure described in Section 2, where the ``24 hours'' period has been selected by $T=\mu
(\tau_{off}+\tau_{on})$, and $\mu$ is another dimensionless parameter corresponding to the average number of pulses emitted during the 24h period. Note that in order to make a reliable DFA analysis we should have at least around ten pulses (see p.301 of Ref.\cite{NEWBOOK} and p.3 of Ref. \cite{MA10}) in each remaining data segment every ``day''. This reflects that even upon 80\% data loss, ten pulses should still remain every ``day'', thus hereafter we select the value $\mu=50$.
Obviously, if we consider a larger $\mu$ value, e.g. $\mu$=100 or 200, our results will become better. 
For $p_r=0.2$, $\lambda=0.1$ and $\mu=50$, we obtain
$p_1=0.45$, $p_2=0.71$ and $p_3=0.83$ (with a plausible estimation
error 5\%). These values change to $p_1=0.57$, $p_2=0.73$ and
$p_3=0.87$ when $p_r$ increases to 0.3 (Table \ref{tab2}).

\subsection{Results from synthetic {noise}.}\label{2.2}

Here, we discuss cases of uncorrelated (in natural time) noise which gives rise to dichotomous signals, e.g. random telegraph signals\cite{NEWBOOK}, which may be confused with SES activities.  In particular, the following three types of synthetic
{noise} will be discussed: (a)Markovian dichotomous
time-series, (b)a noise signal in which the durations of activity
come from a uniform distribution (i.e., $Q_k$ are
{iid} random variables {uniformly distributed in} $[0,2\tau_{on}]$) {or}
(c) from a Gaussian distribution (i.e., $Q_k$ {iid}
originating from a Gaussian distribution with an average duration
$\tau_{on}$ and a standard deviation $\sigma = \beta \tau_{on}$, where
$\beta$ is a dimensionless constant necessarily small, e.g., $\beta=0.1$, to ensure the positivity of the ``durations'' $Q_k$).

{\em (a) Markovian dichotomous signals.} This is the case of the
Random Telegraph Signals (RTS) in which both the durations of
activity and inactivity are {iid} exponential
random variables with average durations $\tau_{on}$ and
$\tau_{off}$, respectively. For the case $p_r=0.2$, $\lambda=0.1$
and $\mu=50$, we obtain $p_1=0.08$, $p_2=0.43$ and $p_3=0.47$.
These values change to $p_1=0.02$, $p_2=0.43$ and $p_3=0.44$ when
$p_r$ becomes 0.3 (Table \ref{tab2}). Hence, we observe that when
using either $p_2$ or $p_3$ it is probable that a Markovian
dichotomous signal may be misinterpreted as SES activity. To avoid
it, we have to perform an additional test described in Appendix A.

{\em (b) Noise signals with uniformly distributed $T_{on}$.} When
the durations $T_{on}(k)$($\equiv Q_k$) are iid uniform random
variables in $[0,2\tau_{on}]$, the results for $p_r=0.2$,
$\lambda=0.1$ and $\mu=50$ are $p_1=0.07$, $p_2=0.33$ and
$p_3=0.37$. These values change to $p_1=0.03$, $p_2=0.29$ and
$p_3=0.31$ when $p_r$ becomes 0.3. Hence, we again observe that
when using either condition (\ref{dfacon}) or (\ref{natcon})
(i.e., the case of $p_3$) for the identification of an SES
activity, it is probable to misinterpret such a noise as SES. To
avoid it, we have to perform an additional test described in Appendix B. 

{\em (c) Noise signals with Gaussian distributed $T_{on}$.} In
such a case for $\beta =0.1$ and $p_r=0.2$, $\lambda=0.1$ and
$\mu=50$, we obtain $p_1=0.08$, $p_2=0.02$ and $p_3=0.10$. These
values change to $p_1=0.01$, $p_2=0.00$ and $p_3=0.02$ when $p_r$
becomes 0.3. These values suggest that it is improbable to
misinterpret such a noise as SES when using  $p_2$ either for
$p_r$=0.2 or $p_r$=0.3. The values of $p_1$ and $p_3$ alone also
achieve a satisfactory distinction between noise and SES
activities but to a lesser extent than $p_2$ (see Table
\ref{tab2}).

\subsection{Results from a real world dataset}\label{2.3}
 {SES activities of appreciably long duration around a few weeks or more, i.e.,
 similar to the one observed in Refs.\cite{UYE02,UYE09} almost two months before the case of Izu
 island swarm in Japan, have not been recorded in Greece. Here, we consider as an example
 the  SES activity  that preceded\cite{NAT09,UYE10} the most recent major EQ in Greece: This, which had
 almost 1.5 days duration, lasted from February 29
 to March 2, 2008, and was followed by a magnitude $M_w$6.4EQ at 38.0$^o$N 21.5$^o$E on June 8, 2008.
 Its original time series, which is not of obvious dichotomous nature, is reproduced in Fig.3(a).
 We now attempt to answer the following question: If such a SES activity would have been recorded in Japan,
 could its identification become possible  by employing the procedure proposed here?}

{Before proceeding we note that the identification
of this SES activity in Greece has been described in detail by Ref.\cite{NAT09}. Chief among the results obtained are the
following: first, the signal as it is evident from an inspection
of Fig.3(a), comprising a number of pulses, is superimposed on a
background which exhibits frequent magnetotelluric (MT)
variations. After subtracting these MT variations\cite{NAT09},
we find the signal depicted in Fig.3(b), which provides the time
series that should be considered for further analysis. Its
analysis in natural time leads to the representation depicted in Fig.4 and the parameters $\kappa_1$, $S$ and $S_-$ resulted in its identification as SES activity.}

In order to answer the question on the possibility to identify this -almost one and a half day long- SES activity after significant data loss and since it may have started any time of the day, a Monte Carlo calculation similar to that described in Section \ref{sec2} was employed. In particular, we randomly select the first segment to keep starting at some time $t_0$ uniformly distributed during the first 24 hours
(i.e, the first
86,400 samples since the sampling rate is 1sample/s), then we discard the experimental data until $t_0+24$h, when the second segment to keep starts -of course, depending on $t_0$ the available SES data may not be enough to allow the selection of a second segment. This is repeated for various values of $t_0$ and the Monte Carlo simulation shows that when
removing 70\% of the data (i.e., $p_r=0.3$) we find a probability
$\approx 67$\% to correctly identify this SES activity ($p_1=0.40$, $p_2=0.54$ and $p_3=0.67$). This
probability becomes somewhat smaller, i.e., $\approx 62$\% upon
increasing the data loss to 80\% ($p_1=0.41$,
$p_2=0.40$ and $p_3=0.62$). These values of the probability are
expected to become markedly larger if the duration of the SES
activity would have been similar to the one observed before the
2000 Izu island seismic swarm in Japan.

\section{{Summary of results and conclusion}}

A flow diagram of the method proposed here to identify in
the remaining data SES activities as well as to avoid to
misinterpret them as noise, is given in Fig. 2. It reveals that
after reading the remaining data in natural time, the DFA exponent
$\alpha$ may be enough to achieve that purpose; otherwise, the
quantities $\kappa_1$, $S$, and $S_-$ as well the statistics of the
duration of pulses should be also studied.

Our main conclusion states that when employing two modern
techniques, i.e., natural time analysis and DFA, a distinction
between SES activities (critical dynamics) and artificial
noise becomes possible even after removing periodically
a significant portion of the data. In particular{,
when using {\em synthetic} long duration SES activities, upon} removing
70\% of the data (e.g. the data from 06:00 to 22:00 LT like in Japan),
we have a probability around 87\%, or larger, to identify
correctly an SES activity. This probability becomes somewhat
smaller, i.e., 83\%, when the data removal increases to 80\%.
{In addition, when investigating a real world data
set referring to an SES activity of only one and a half day
duration, the following results are obtained: upon removing 70\%
(80\%) of the data, we find a probability $\approx$67\%(62\%) to
correctly identify this SES activity. The probability is expected
to increase significantly for SES activities of longer duration,
i.e., a few weeks or more as in the case of the Izu island seismic
swarm.}

\appendix

\section{An additional test to discriminate Markovian dichotomous signals from SES activities}
Since Markovian dichotomous signals are probable to be misinterpreted as SES activities because their $p_3$ values are close to 50\%, we
additionally check the ratio $\sigma (T_{on})/\tau_{on}$ of
the standard deviation $\sigma (T_{on})$ over the average duration
of the pulses $\tau_{on}$. The Markovian dichotomous signals have
a ratio varying (even in the case of 50 pulses) in the range
[0.72,1.34] with probability 98\%. For the SES model used in subsection \ref{2.1},
the probability to have a ratio larger than 0.70 is only 0.2\%. In
other words, we should compare the
$\sigma(T_{on})/\mu(T_{on})$-value of the signal under
investigation with the corresponding value of the exponential
distribution having a number of samples equal to the number of pulses in the signal. Thus,
when using either condition (\ref{dfacon}) or (\ref{natcon})
(i.e., the case of $p_3$) for the identification of an SES
activity, we {\it should} also study the statistics of the
durations of the recorded pulses to avoid misinterpreting a
Markovian dichotomous signals as SES.

\section{An additional test to discriminate noise signals with uniformly distributed $T_{on}$ from SES activities}
As mentioned in subsection \ref{2.2}(b), $p_3$ in this case may reach values around 33\%. In order to avoid misinterpreting such a signal as SES activity, we again (see Appendix A) 
 resort  to the statistics of the durations of
the pulses. In particular, we consider the ratio $\rho$ of the
range of durations for such a noise over their standard deviation,
i.e., $\left[(T_{on})_{max}-(T_{on})_{min}\right]/\sigma
(T_{on})$($\equiv \rho$), which varies  (even in the case of 50
pulses)  in the range [3.0,3.8] with probability 95\%. Moreover,
the probability $P(\rho < 3.7)$ that $\rho$ is
smaller than 3.7 is around 94\%.  On the other hand, the
probability that an SES activity, modelled according to subection \ref{2.1}, has
a $\rho$-value larger than 3.8 is about 87.5\% and increases to
92\% when we consider a $\rho$ value larger than 3.7. Thus, when
using either condition (\ref{dfacon}) or (\ref{natcon}) (i.e., the
case of $p_3$) for the identification of an SES activity, we {\it
should} also study the statistics of the durations of the recorded
pulses to obtain $\rho$. Then there is approximately 90\%
probability to safely distinguish SES from a noise signal with
$T_{on}$ uniformly distributed even in the extreme case of
recording only 50 pulses. This result becomes better if 100 pulses
are recorded in the noise-free segments. In this case, $\rho$
varies in the range [3.1,3.8] with probability 98\%, and $P(\rho <
3.7)$ is 96\%; on the other hand, the synthetic SES activities
treated in subsection \ref{2.1} result in a small $P(\rho<3.7)$ value $\sim$3\%.

%\section*{References}
%\bibliographystyle{apsrev}
%\bibliography{refestecto}

\begin{figure}
\noindent\includegraphics[width=20pc]{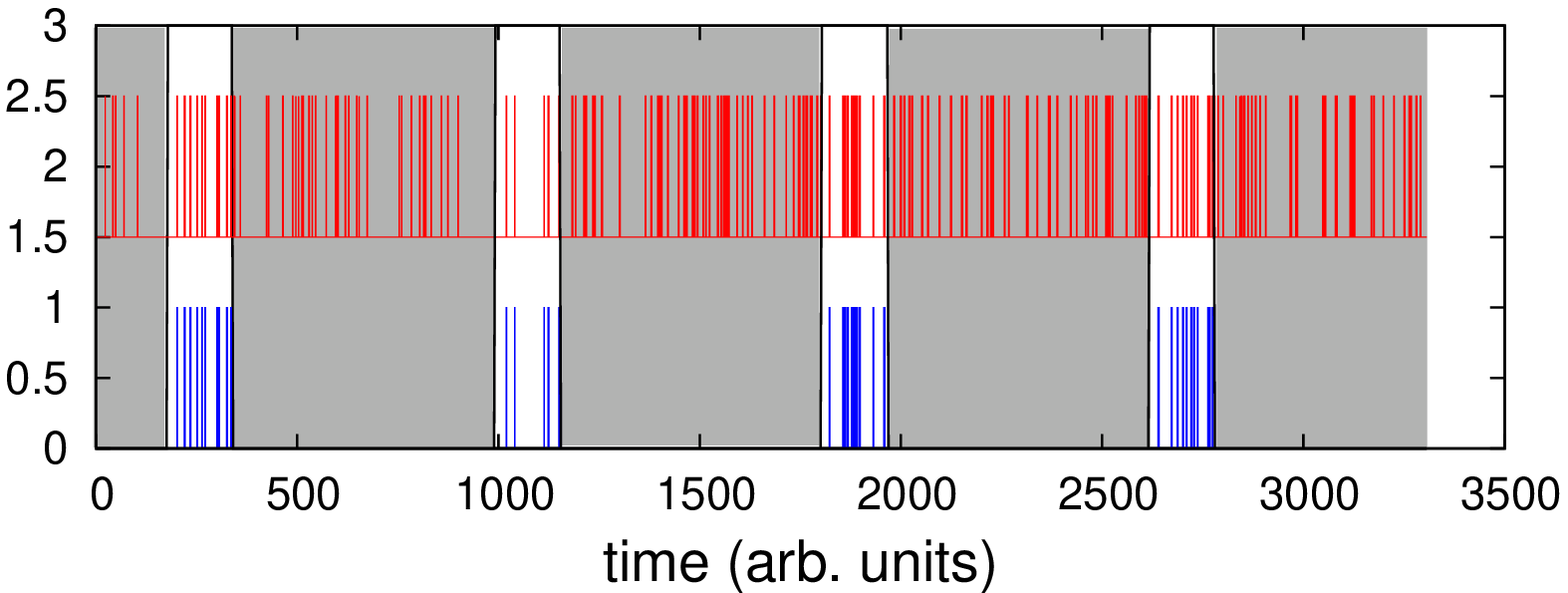} \caption{An example of the procedure described in subsection \ref{2.1}, showing data segments that are periodically removed every T$\approx$813 arbitrary units from the original dichotomous time-series (red). The grey shaded areas correspond to the high noise periods (e.g. 06:00-22:00 LT in Japan) which have to be discarded daily.}
\end{figure}

\begin{figure}
\noindent\includegraphics[width=20pc]{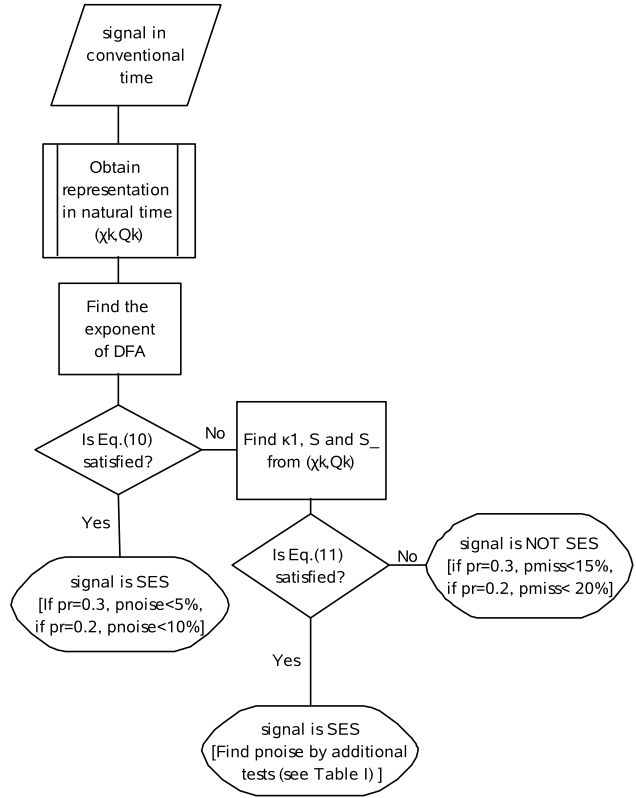}
\caption{Flow-chart of the method proposed.}
\end{figure}

\begin{figure}
\noindent\includegraphics[width=20pc]{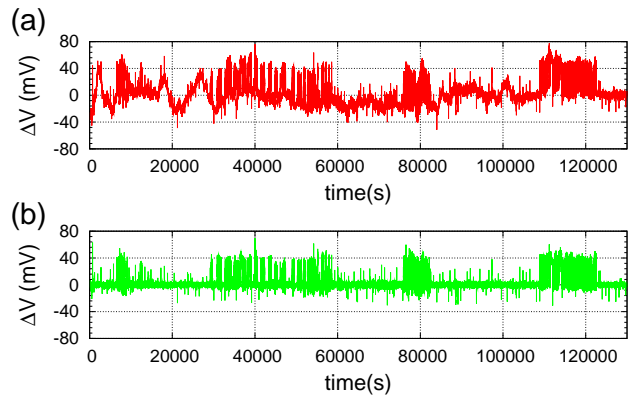}
{\caption{The SES activity from February 29 to
March 2, 2008: (a)original time-series (b) the same as (a) but
after subtracting the MT background variations by applying the
procedure described in Ref.\onlinecite{NAT09}.}}
\end{figure}

\begin{figure}
\noindent\includegraphics[width=30pc]{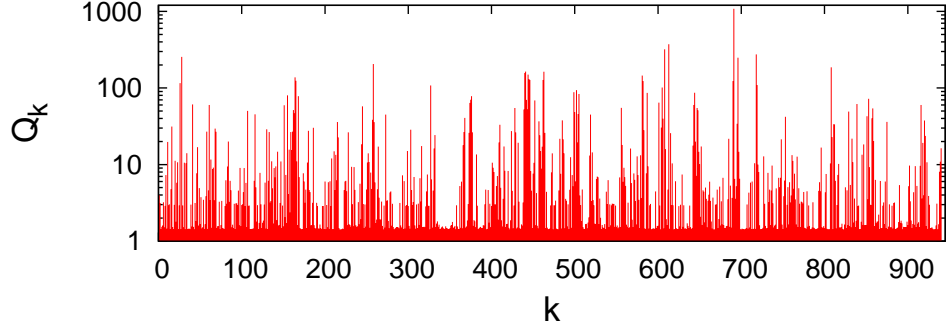}
{\caption{Natural time representation of the signal depicted in Fig.3(b) according to Fig.7 of Ref. \onlinecite{NAT09}. The ``durations'' $Q_k$ are here presented in arbitrary scale.}}
\end{figure}

\begin{table}
\caption{ The values of $p_1$, $p_2$ and $p_3$ (with an estimation
error 5\%) obtained with $\lambda=0.1$ and $\mu=50$ for the
synthetic SES activities (SES), Markovian dichotomous signals
(Markovian), noise signals with uniformly distributed $T_{on}$
(Uniform) and noise signals with Gaussian distributed $T_{on}$
with $\beta =0.1$ (Gaussian). The two cases $p_r=0.2$ or $p_r=0.3$
are shown.} \label{tab2}
%\begin{ruledtabular}
%\begin{minipage}
\begin{tabular}{ccccc}
\hline
 Type of signal & $p_r$ & $p_1$(\%) &  $p_2$ (\%) & $p_3$  (\%) \\
\hline
SES & 0.2 & 45  & 71      &   83  \\
&  &  &       &    \\
Markovian\footnotemark[1] & 0.2 & 8 & 43 &   47       \\
Uniform\footnotemark[2] & 0.2 & 7 &  33 &  37      \\
Gaussian & 0.2 & 8 & 2 &  10     \\
 &  &   &    &    \\
\hline
SES & 0.3 & 57  & 73      &   87  \\
 &  &   &    &    \\
Markovian\footnotemark[1] & 0.3 & 2 & 43 &   44       \\
Uniform\footnotemark[2] & 0.3 & 3 & 29 &  31      \\
Gaussian & 0.3 & 1 & 0 &  2 \\
 &  &  &     &    \\
\hline
\end{tabular}

\footnotemark[1]{Secure distinction from SES is achieved by the
statistics of $\sigma(T_{on})/\mu(T_{on})$.}

\footnotemark[2]{Secure distinction from SES is achieved by the
statistics of
$\rho=\left[(T_{on})_{max}-(T_{on})_{min}\right]/\sigma
(T_{on})$, see Appendix 2.}
 %\tablenotetext{3}{The distinction between such noises and SES can be further elaborated by applying the Kolmogorov-Smirnov test.}
%\end{minipage}
%\end{ruledtabular}

\end{table}

\end{document}